\documentclass[iop]{emulateapj}

\usepackage[english]{babel}
\usepackage[utf8x]{inputenc}
\usepackage[T1]{fontenc}

\usepackage{amsmath,bm}
\usepackage{amssymb}
\usepackage{graphicx}
\usepackage[colorlinks=true, allcolors=blue]{hyperref}
\usepackage{booktabs}
\usepackage{ulem}
\usepackage{color}
\usepackage{amsmath}
\usepackage[symbol]{footmisc}
\usepackage{natbib}
\usepackage{lineno}

\shorttitle{Efficiency of Planetesimal Accretion - III: Planetesimals Beyond Saturn}
\shortauthors{Haghighipour, Podolak, Podolak}

\begin{document}

\title{Detailed Calculations of the Efficiency of Planetesimal Accretion in the Core-Accretion Model -III:
The Contribution of Planetesimals Beyond Saturn}

\author{Nader Haghighipour\altaffilmark{1,2,3}, Morris Podolak\altaffilmark{4}, and Esther Podolak\altaffilmark{5}}

\altaffiltext{1}{Planetary Science Institute, Tucson, AZ 85719, USA} 
\altaffiltext{2}{Institute for Astronomy, University of Hawaii-Manoa, Honolulu, HI 96822, USA}
\altaffiltext{3}{Institute for Advanced Planetary Astrophysics, Honolulu, HI, USA}
\altaffiltext{4}{Department of Geosciences, Tel Aviv University, Tel Aviv, Israel 69978}
\altaffiltext{5}{Liacom, Holon, Israel}

\begin{abstract}
Continuing our initiative on advancing the calculations of planetesimal accretion in the core-accretion model, 
we present here the results of our recent study of the contributions of planetesimals around and beyond the orbit 
of Saturn. In our first two papers [ApJ, 899:45; 941:117], where our focus was on the effects of the Sun and Saturn, 
the initial distribution of planetesimals was limited to the regions around the accretion zone of a growing Jupiter. 
In this paper, we expanded that distribution to regions beyond the accretion zone of Saturn. We integrated the orbits 
of a large ensemble of planetesimals and studied the rate of their capture by the growing proto-Jupiter. In order to 
be consistent with our previous studies, we did not consider the effect of the nebular gas. Results demonstrated that 
the exterior planetesimals, especially those beyond 8 AU, have only slight contributions to the growth and metallicity 
of the growing Jupiter. The final mass and composition of this planet is mainly due to the planetesimals inside and 
around its accretion zone. Our study shows that although the rate of capture varies slightly by the size and composition 
of planetesimals, in general, the final results are independent of the size and material of these bodies. Results also 
pointed to a new finding: the rate of accretion follows the same trend as that of the evolution of Jupiter's envelope, 
with the largest accretion occurring during the envelope's collapse. We present details of our analysis and discuss 
the implications of their results.
\end{abstract}

\section{Introduction}
Planetesimal accretion is fundamental to planet formation and the chemical composition of planetary bodies.
In the core-accretion model of giant planet formation, the accretion of planetesimals results in 
the formation of the core while the interaction of these bodies with the proto-giant planet's envelope, 
both during gas-accretion and at the stage of the envelope-collapse, contributes to the metallicity of 
the envelope and the luminosity of the resulting giant planet. It also determines the onset of the 
collapse of the envelope.

The outcome of planetesimal-envelope interaction varies based on the internal 
properties of the planetesimal, the state of the envelope, and the planetesimal-envelope encounter velocity.
For instance, the efficiency of the vaporization of a planetesimal due to its heating by gas drag, and the
subsequent deposition of its chemical compounds are strong functions of the planetesimal's
material strength and impact velocity. They also depend heavily on the
local density and temperature of the envelope (i.e., its internal structure) which determine its
response as the planetesimal travels through the gas.

These intertwined connections between the growth and composition of a giant planet, the dynamical state
of its envelope, and the composition and dynamical properties of planetesimals strongly imply that
any model of the formation of giant planets needs to be able to take these interactions into account, both accurately and 
self-consistently. This paper is the third of a series where we present the details of such a model.

We began our initiative by exploring the effects of different processes in a methodical way. In our first paper, 
\citep[][hereafter, Paper-I]{Podolak20}, we included only the gravitational force of the Sun as this effect had been 
neglected in the original calculations of the core accretion model \citep{Pollack96}. In that paper, we also
introduced our special purpose integrator, ESSTI (Explicit Solar System Trajectory Integrator) which has been designed
for the specific purpose of carrying out orbital integrations while calculating planetesimal-envelope interactions.
In our second paper, \citep[][hereafter, Paper-II]{Haghighipour22}, we advanced our calculations by including 
the perturbation of a second planet in the orbit of Saturn. We demonstrated that the gravitational perturbation of an 
exterior planet reduces the rate of planetesimal accretion by scattering many planetesimals that are interior to its orbit, 
to distances outside the proto-giant planet's accretion zone. 

Because the purpose of Paper II was to demonstrate the effects of an external giant planet, the distribution of planetesimals
was limited to orbits interior to this body. While the results presented in that paper stand for such distributions, there is a 
chance that the perturbing effect of the outer planet may be countered (at least, partially) if planetesimals existed exterior to its 
orbit as well. In this study, we aim to explore this scenario.

\begin{figure}[ht]
\vskip -24pt
\hskip -25pt
\includegraphics[scale=0.41]{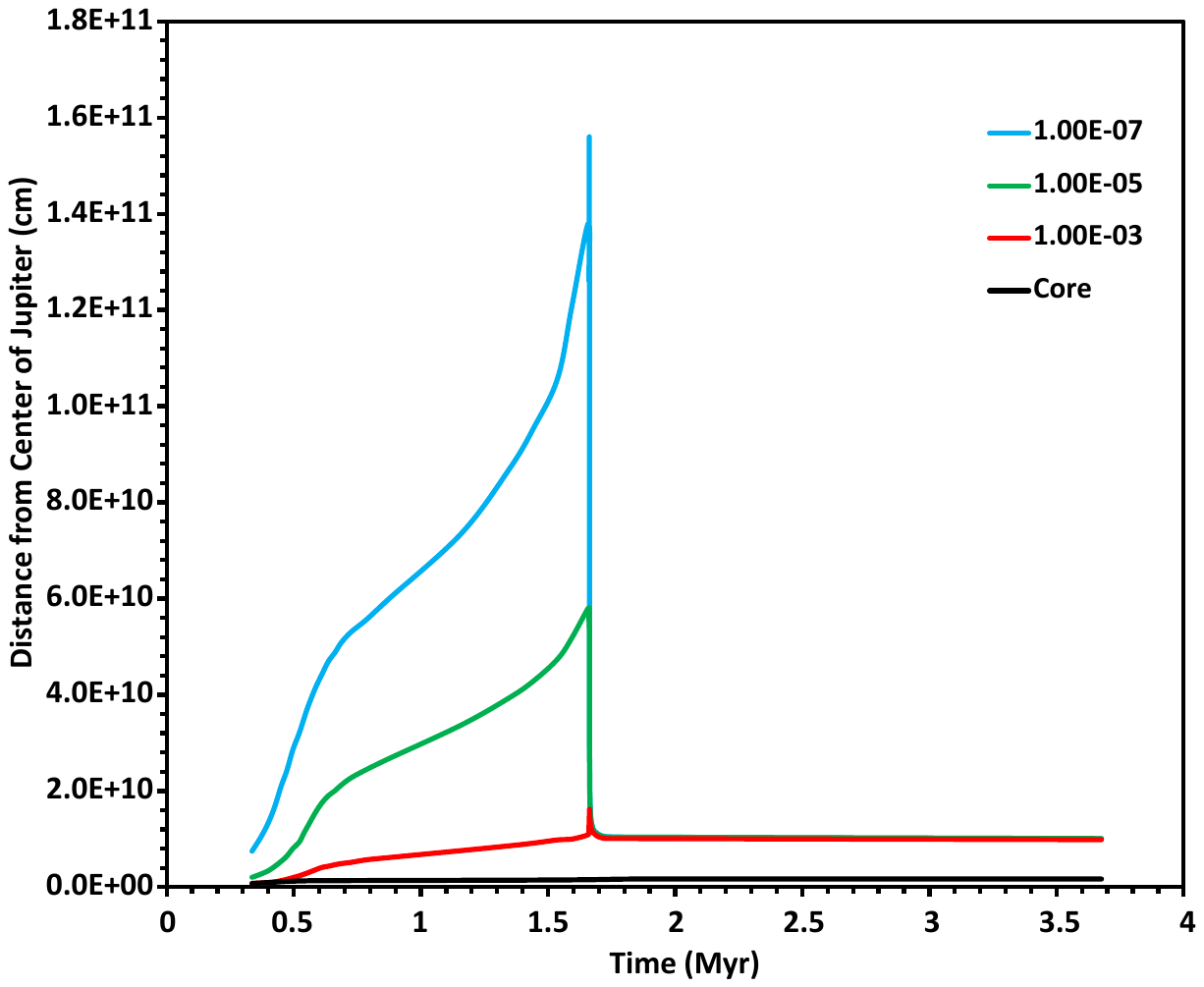}
\vskip -45pt
\hskip -23pt
\includegraphics[scale=0.36]{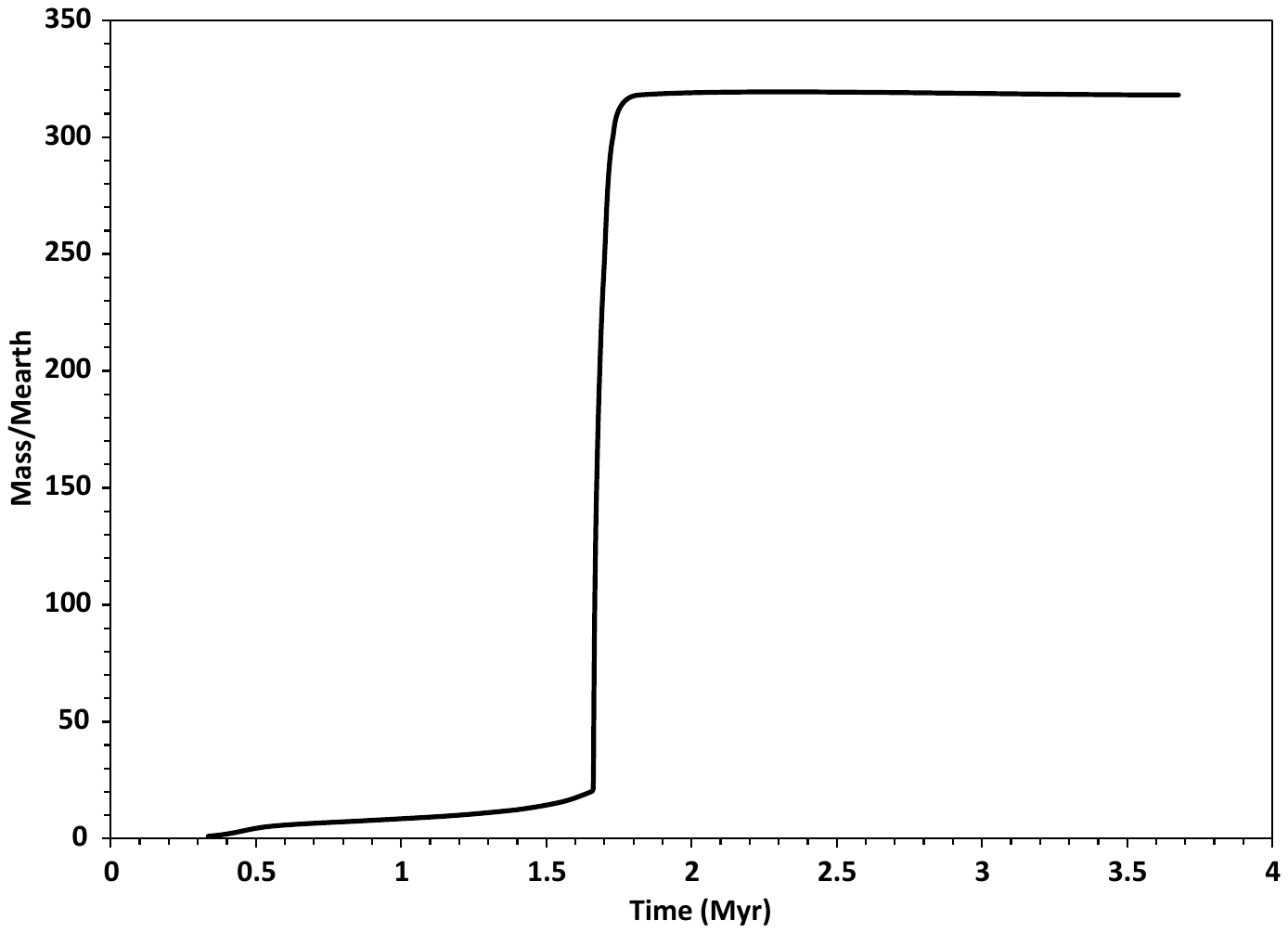}
\vskip -30pt
\caption{Top: The graphs of the time-varations of the density of the envelope at different distances from the center of 
the proto-giant planet based on the models described by \citet{Lozovsky17}. Shown are gas densities of 
${10^{-7}}\, {\rm g \, {cm^{-3}}}$ (blue line), ${10^{-5}}\, {\rm g \, {cm^{-3}}}$ (green line), 
and ${10^{-3}}\, {\rm g \, {cm^{-3}}}$ (red line). The radius of the core is shown by black line.
Bottom: The graph of the time-variation of the total mass of the planet. Note the runaway gas accretion
and collapse of the envelope at 1.7 Myr.}
\label{fig1}
\end{figure}

In structuring this paper, we follow the same format as in Papers I and II. We begin by introducing our system and 
explaining our numerical approach. We will then continue by presenting our results and their analysis. To
avoid repetition, we refrain from lengthy explanations and refer the reader to Papers I and II for historic background,
results of previous studies, and for more details on the physics of planetesimal-envelope interaction
and our integrator, ESSTI.

\section{The Initial Set up and Numerical Approach}

\subsection{The System}
Our system consists of the Sun, a proto-giant planet in the orbit of Jupiter, a giant planet in the orbit
of Saturn, and a mass-less planetesimal. We consider the proto-giant planet to be an extended mass and, for the 
mere sake of simplicity, we assume that it has a fully condensed core with a well-defined 
surface that separates heavy elements and H/He. Because we are interested in the contributions of planetesimals exterior 
to the orbit of the outer planet, we consider the semimajor axes of the planetesimals to vary randomly between 8 AU to 12 AU.
To avoid computational complications, we do not consider the growth of the outer planet through accretion of planetesimals. 
Instead, we assume that the planet is fully formed and include its gravitational effect as a constant perturbation. Also,
we do not consider the effect of the nebular gas. Although in 
reality, the drag force of the nebula affects the dynamics of planetesimals and, therefore, their encounter
velocities with the giant planet's envelope, for the purpose of this study, its inclusion is unnecessary as our goal
is to merely demonstrate the proof of concept. 

The variations of the radius, mass and internal density of the proto-giant planet are included
using model A of \citet{Lozovsky17}. In this model, which has been developed for the core-accretion scenario, the growth
of a planet is followed in a background solid material of 100 km planetesimals with a surface density of 
$\sigma=6\, {\rm {g \, cm^{-2}}}$. The top panel of
figure 1 shows how the radius of a constant density surface in the envelope varies with time as the planet grows and evolves. 
The blue line represents
the gas density of ${10^{-7}}\,{\rm g \, cm^{-3}}$, the green and red lines correspond to the gas densities of 
${10^{-5}}\,{\rm g \, cm^{-3}}$ and ${10^{-3}}\,{\rm g \, cm^{-3}}$, respectively, and the black line shows the growth of
the core. The bottom panel shows the variation of the total mass of the planet with time. 
As shown here, as the proto-giant planet grows, the mass and radius 
of its envelope grow as well. As a result, the density of the gas varies with time at different elevations inside the envelope. 
In this model, at approximately 1.7 Myr, the envelope becomes unstable and the runaway gas-accretion ensues. 

To assign a composition to the planetesimals, we note that in the region of the formation of gas-giant planets, the 
vast majority of planetesimals are a mixture of ice and rock. We, therefore, consider such a mix composition and 
following the trend observed in the composition of large bodies in the outer solar system, we assume a mixture of 
30\% ice and 70\% rock with a bulk density of $2\,{\rm g \, cm^{-3}}$, similar to that of Pluto, Titan, Triton.
 
It is important to note that because in a mix-composition planetesimal, the strength and extent of the planetesimal's 
response is mainly driven by the efficiency of the evaporation of its ice (see section 3.2 in Paper II), 
it would be unnecessary to explore a large range of ice-to-rock ratios. 
For that reason and for the sake of completeness, we also consider compositions at the low and 
high ends of the spectrum and include  planetesimals of pure ice with a bulk density of $1\,{\rm g \, cm^{-3}}$ 
and pure rock with a bulk density of ${3.4}\,{\rm g \, cm^{-3}}$. For each of these three compositions, we run 
simulation for planetesimals with 1 km, 10 km, and 100 km radii.

\subsection{Numerical Integrations}

We integrated 2005 different cases of our four-body system where we assigned random small eccentricity and zero 
inclination to the planetesimal's orbit. Integrations were carried out for 3 Myr, approximately twice  the time that 
it takes for the envelope of the proto-giant planet to collapse. Each integration started
with the planetesimal at its initial semimajor axis between 8 AU and 12 AU. We integrated the system using 
the hybrid routine in the $N$-body integration package Mercury6 \citep{Chambers99} until
the integrator identified a collision between the planetesimal and the giant planet. At that time, we stopped the
integration and passed the orbital elements of all bodies to our special purpose integrator ESSTI. 
Integration was then continued using a $4^{\rm th}$ order Runge-Kutta integrator.
During the integration, the evolution of the mass 
of the proto-giant planet was included following the prescription by \citet{Lozovsky17}.
As mentioned before, ESSTI includes all physical processes relevant to the planetesimal-envelope interactions.
We refer the reader to sections 2.3 and 2.4 of Paper-II for more details on the 
above processes and our ESSTI integrator. We integrated our systems 
for different combinations of the size and composition of the planetesimals (see section 2.1) and for each of those 
combinations, we considered three different masses for the planet in the orbit of Saturn: Full Saturn mass, 1/3, 
and 1/10 of the mass of Saturn.

\section{Results and Analysis}

During an integration, the orbital evolution of a planetesimal manifests itself through one of 
the following four scenarios: Either the planetesimal 1) enters the proto-giant planet's envelope, 2) collides 
with the planet in the orbit of Saturn, 3) is scattered to a larger orbit outside the integration region, or 
4) stays in the system for the duration of the integration. Results of our integrations have demonstrated 
that the first three scenarios are the most probable outcomes. In fact, given the strong perturbation 
of the two giant planets, it is expected that any planetesimal that maintains its orbit
for the 3 Myr of the integration will be eventually scattered out. A small fraction of our systems
($\sim 5\%$ to 20\%, depending on the assumed mass of the outer planet) do in fact show this behavior.
Figure 2 shows a sample of these systems for a 100 km planetesimal 
and for the highest and lowest masses of the outer planet. Here, the initial and final eccentricity of the planetesimal 
(black and red, respectively) are shown in terms of its initial semimajor axis.
As shown by red circles, in almost all cases, the orbit of the planetesimal becomes highly eccentric indicating that, 
had the integration been continued, it would have been ejected from the system. 

When a planetesimal enters the envelope, it loses mass due to the evaporation caused by its interaction with the envelope's gas.
This mass loss may be partial, meaning that, after depositing some of its material, the planetesimal may exit the envelope and 
continue its orbit outside the proto-giant planet. Or, the mass-loss may be complete, resulting in the full capture of 
the body. In that case, either the entire mass of the planetesimal is deposited in the envelope, or as the planetesimal
penetrates deeper inside the proto-planet, it maintains some of its mass until it collides with the planet's core
and contributes to the core's growth. Figure 3 shows the fraction (in percentage) of the initial mass of 
the planetesimals in the region of 8 AU to 12 AU that was accreted by the proto-giant planet through the above two processes.
To demonstrate the effect of the planetesimal's size, the figure shows the results for a 1 km and a 100 km body. 
The colors represent ice (blue), rock (gray) and mix-composition (orange) for a 100 km planetesimal 
and ice (purple), rock (red), and mix-composition (green) for a 1 km object. From top to bottom, the panels show the
results for the mass of the outer planet being 1/10, 1/3, and full mass of Saturn. 
As shown here, as the mass of the outer planet increases, the fraction of the accreted material becomes smaller.
Similar results are shown in figure 4 for a 100 km pure rock and a 1 km pure ice. As shown by this figure, an increase in the mass
of the outer planet from 1/10 to full Saturn-mass results in a decrease of almost 8\% in the accreted material. 
This decrease is due to the fact that for large values of the mass of the outer planet, the perturbation of this
body ejects more planetesimals out of the system resulting in fewer bodies entering the 
proto-planet's envelope. Figure 5 shows this for all cases of the figure 3. As demonstrated by this figure, increasing the 
mass of the outer planet causes a significant number of planetesimals to be scattered out of the system. For instance,
when the mass of this planet changes from 1/10 of Saturn to the full Saturn-mass, the rate of the ejected planetesimals 
increases by more than two folds from $\sim 12\%$ to almost 27\%.

\begin{figure}[ht]
\vskip -15pt
\hskip -30pt
\includegraphics[scale=0.38]{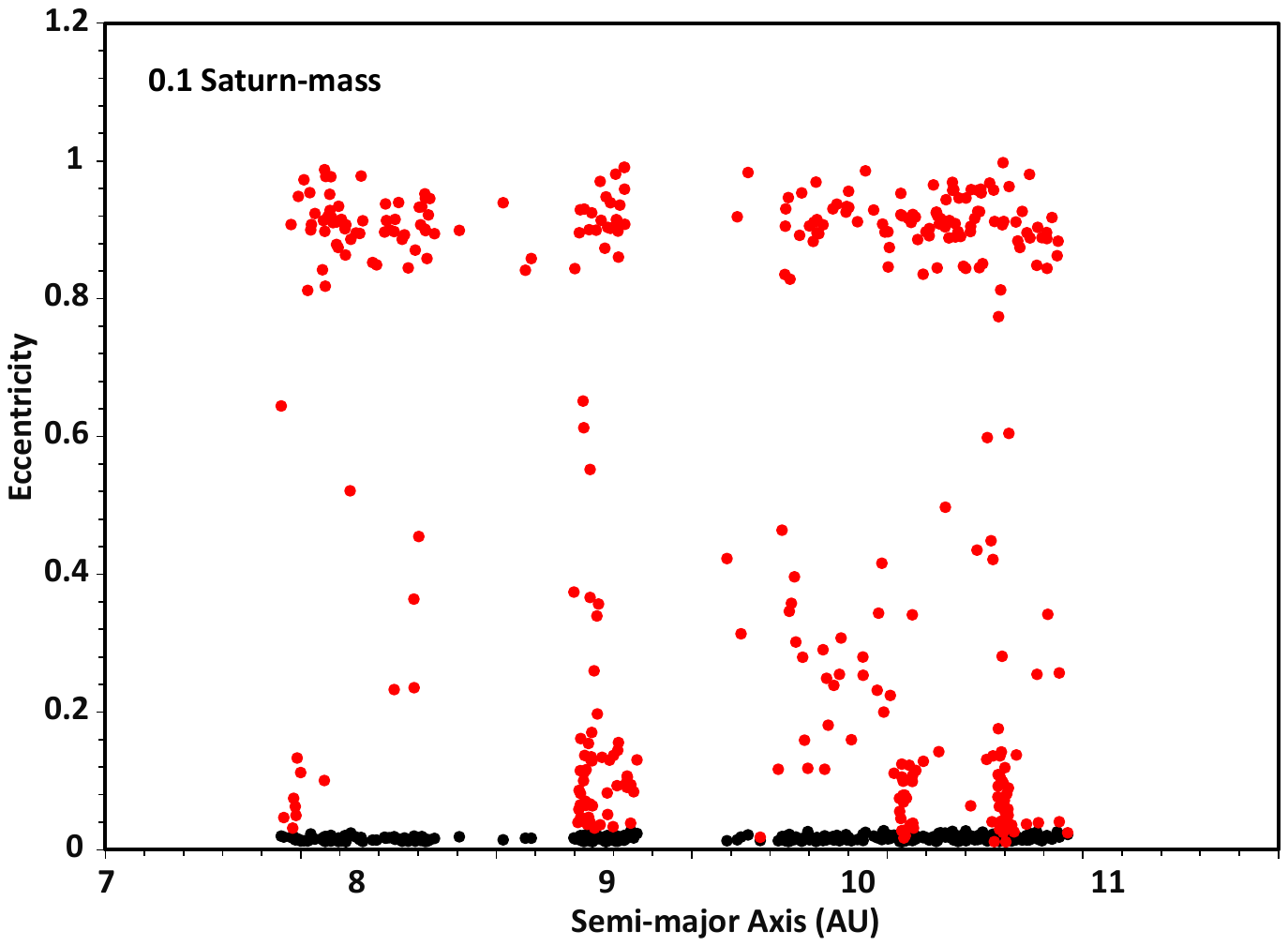}
\vskip -45pt
\hskip -30pt
\includegraphics[scale=0.38]{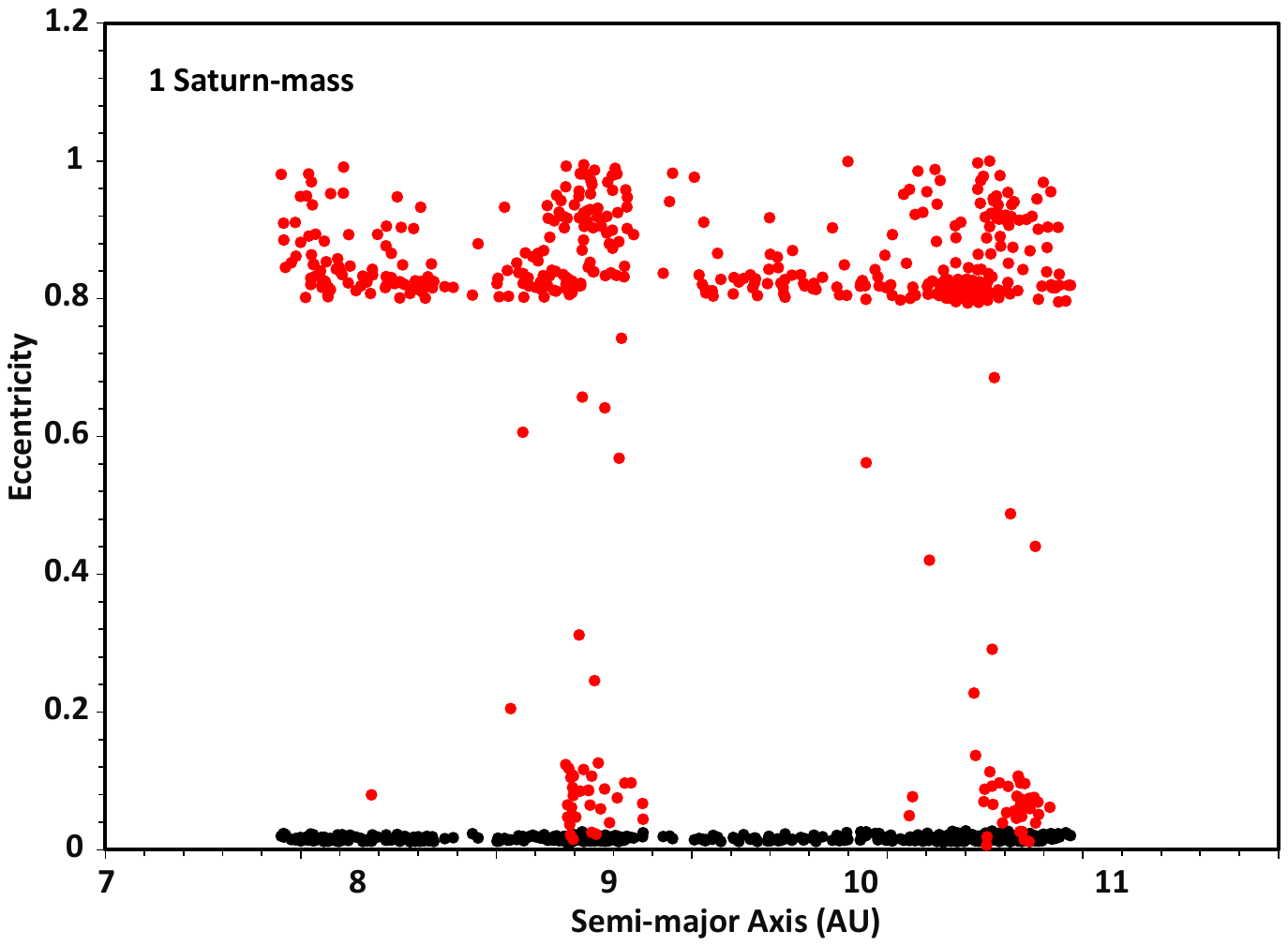}
\vskip -30pt
\caption{Graphs of the final eccentricity of the planetesimal (red dots) in terms of its initial seimmajor axis. 
The black dots show the planetesimal's initial eccentricity. The top panel repsents the results for the mass of 
the exterior planet equal to 0.1 Saturn and the lower panel corresponds to the full Saturn-mass.}
\label{fig2}
\end{figure}

An interesting result depicted by figure 3 is the appearance of a trend in the time evolution of 
the accreted material. As shown by this figure, for all values of the mass of the outer planet as well as the size and 
composition of the planetesimal, the rate of accretion increases gradually until the moment 
of the rapid contraction of the envelope. At that time, accretion becomes very rapid and when the proto-giant planet 
reaches the hydrostatic equilibrium, the accretion rate stabilizes. The strongest contributions 
occur during the collapse of the envelope. During this time, the enhancement in the local density of the envelope 
strengthens the interaction of the planetesimal with the gas increasing its vaporization efficacy. As a result, more of the mass
of the planetesimal is deposited in the envelope.  
Our integrations demonstrated that for a 1/10 Saturn-mass planet, from all the
accreted material, approximately 20\% occurred before the envelope collapse and 80\% during and after that period. 
For a 1/3 Saturn-mass planet, these values became 50\% and 50\% and stayed roughly the same for a full Saturn-mass body.

\begin{figure}[ht]
\vskip -15pt
\hskip -20pt
\includegraphics[scale=0.41]{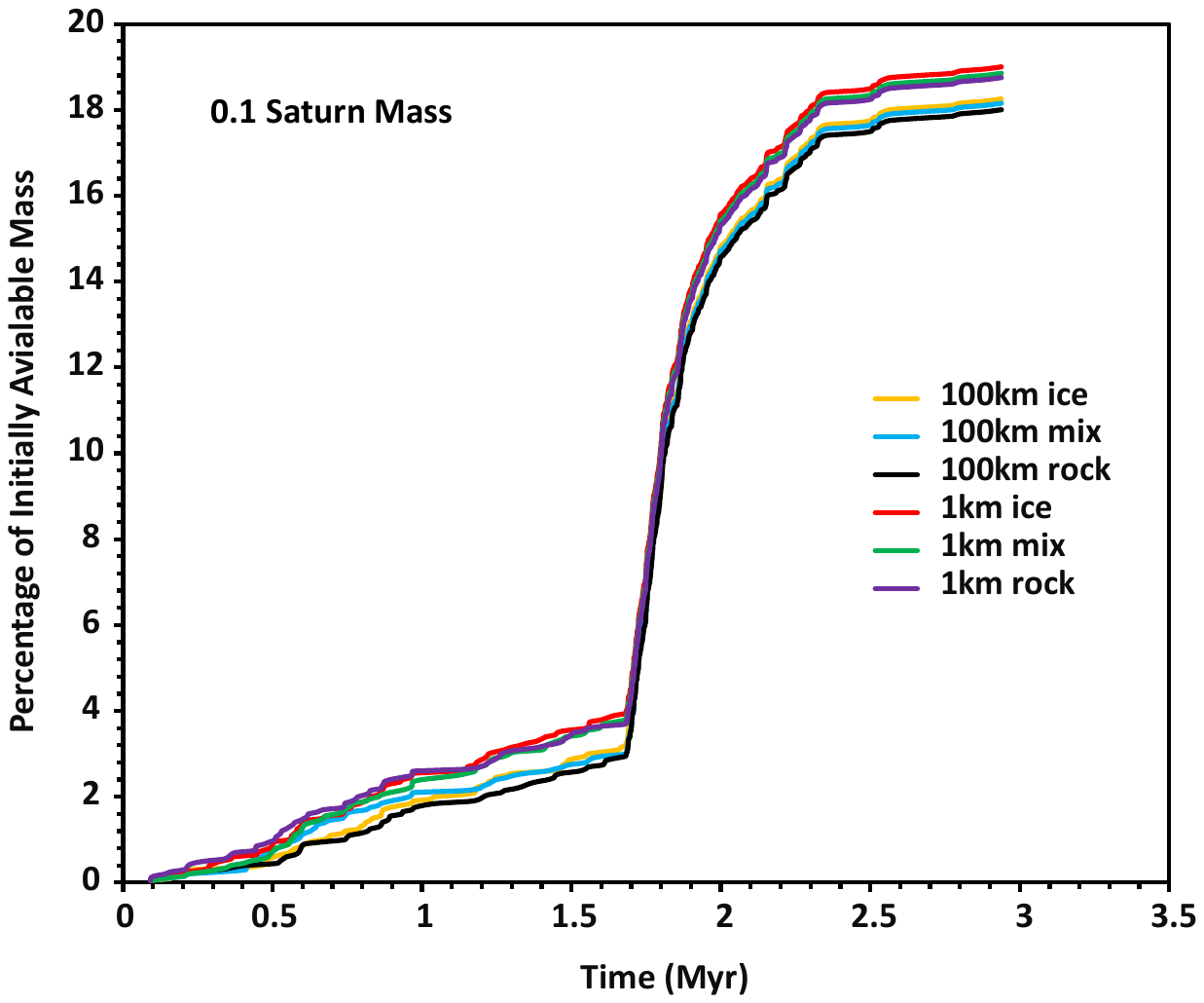}
\vskip -50pt
\hskip -20pt
\includegraphics[scale=0.41]{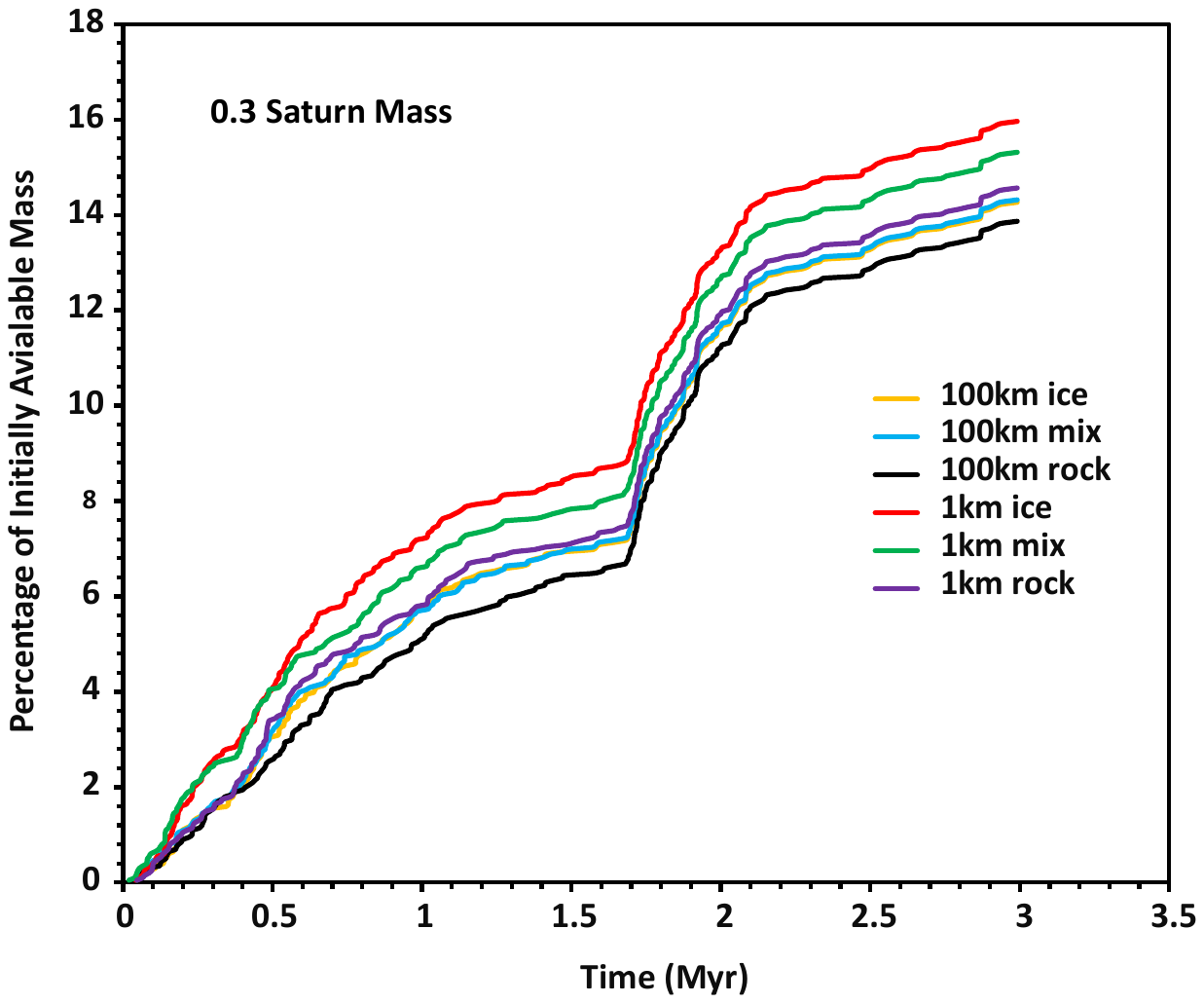}
\vskip -50pt
\hskip -20pt
\includegraphics[scale=0.41]{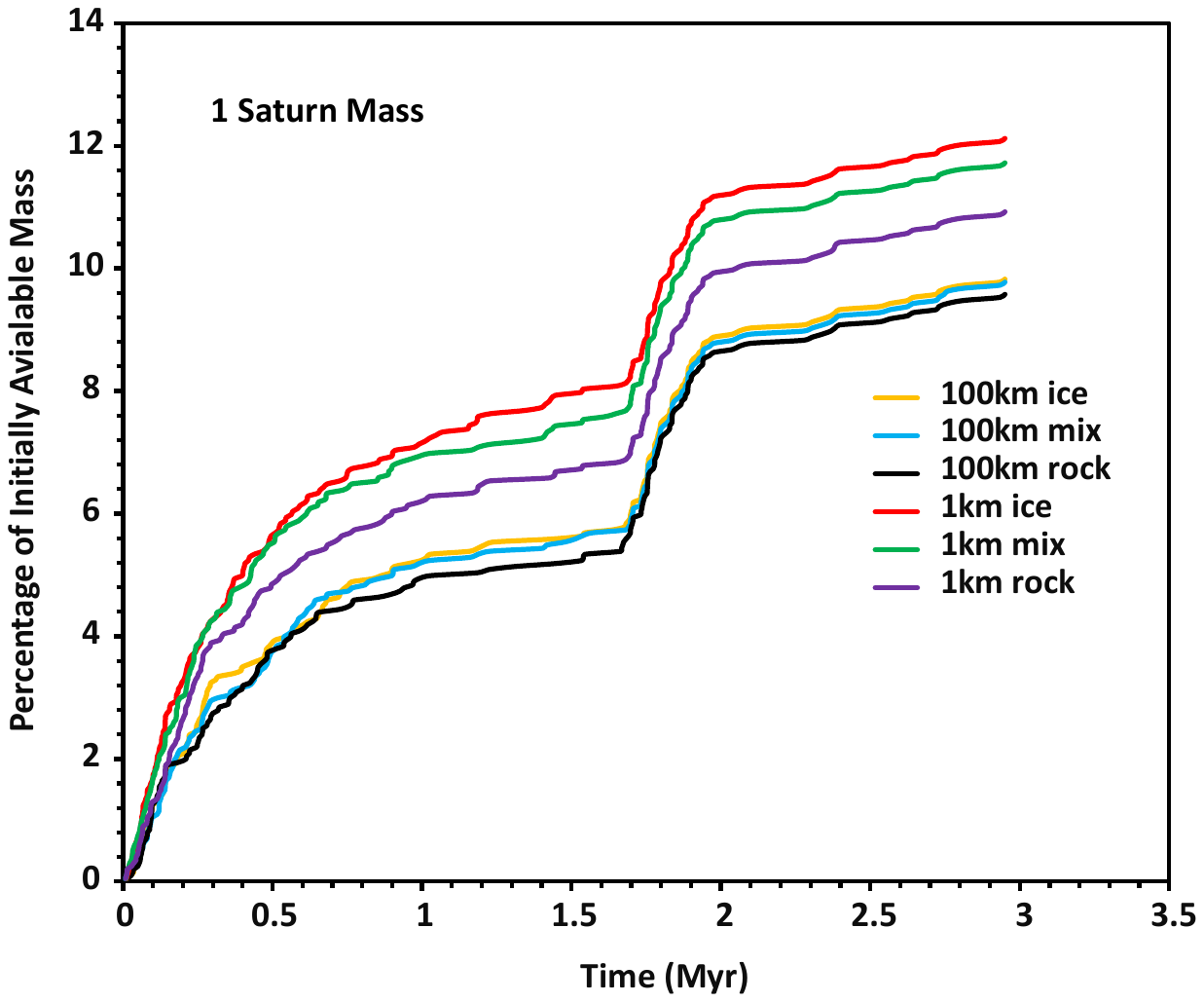}
\vskip -25pt
\caption{Graphs of the accreted planetesimal-mass as a percentage of the initially available mass in the region 
of 8 AU to 12 AU. Each panel shows the results for a 1 km and 100 km ice, rock and the mix planetesimal.
From top to bottom, the panels correspond to three different values of the mass of the outer planet.}
\label{fig3}
\end{figure}

\begin{figure}[ht]
\vskip -15pt
\hskip -20pt
\includegraphics[scale=0.41]{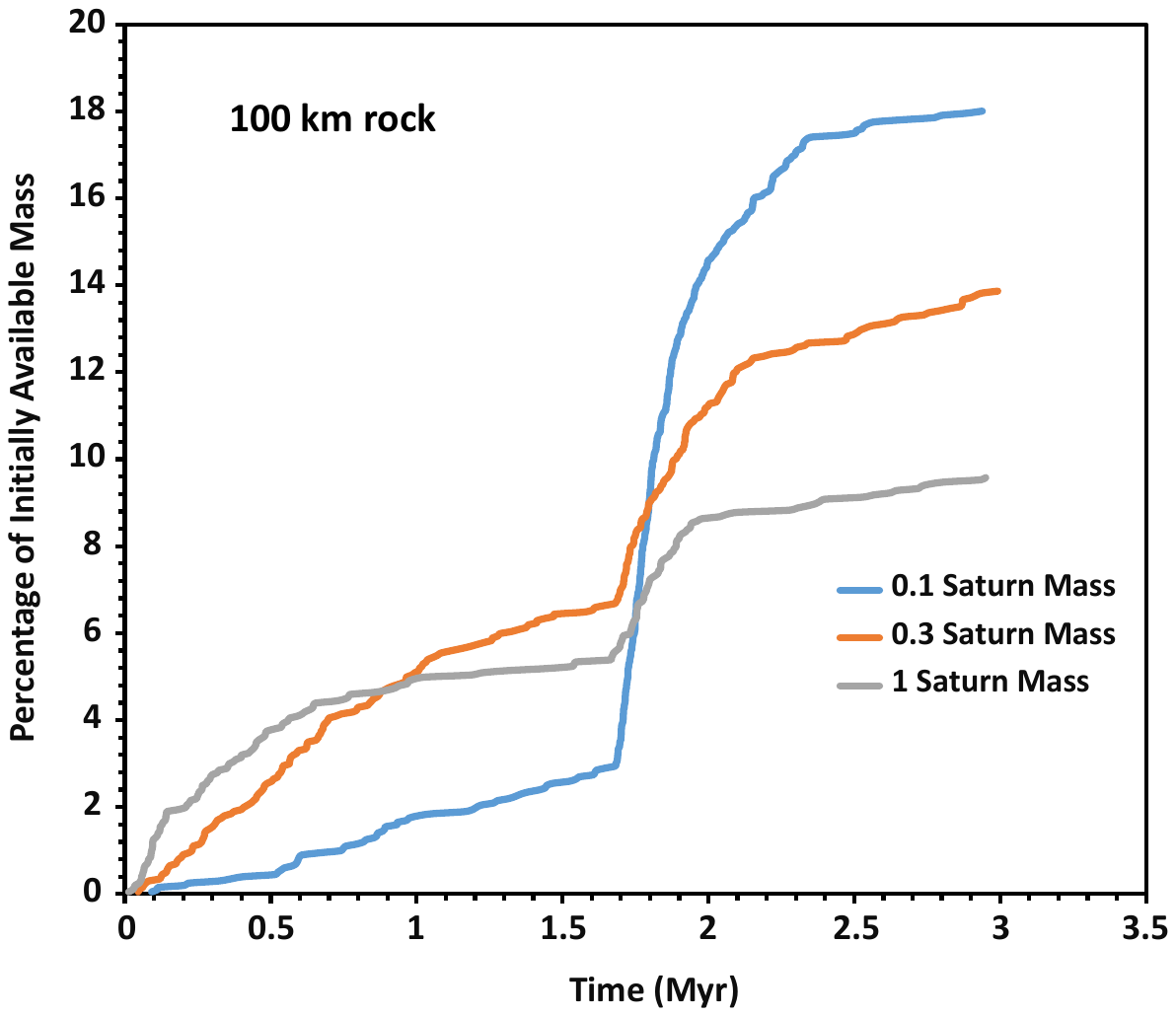}
\vskip -45pt
\hskip -20pt
\includegraphics[scale=0.41]{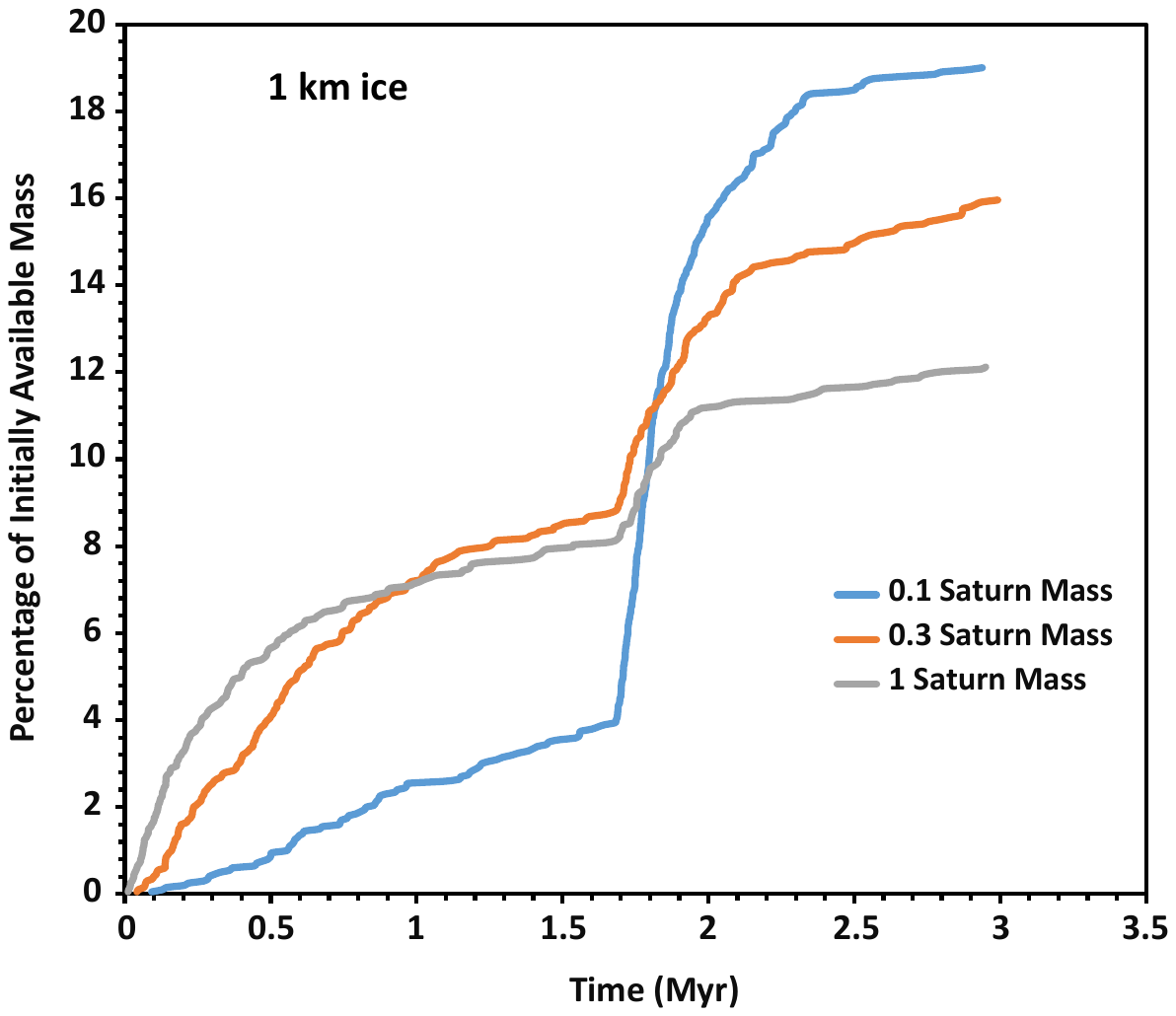}
\vskip -25pt
\caption{Graphs of the accreted planetesimal-mass as a percentage of the initially available mass in the region 
of 8 AU to 12 AU for a 100 km rock and 1 km ice. Each figure shows the results for three different values
of the mass of the outer planet.}
\label{fig4}
\end{figure}

Figure 3 also shows that for a given mass of the outer planet, the above trend is almost independent of the 
size and composition of the planetesimal. For a 1/10 Saturn-mass, the difference between the accreted material 
for different compositions and different planetesimal sizes are negligible. For larger values of the mass of the 
planet, a small spread appears for different compositions of 1 km planetesimals, whereas for the 100 km 
planetesimals, the accretion rates stay very close together. The appearance of the spread in the the accretion
of 1 km bodies is due to the fact that as the perturbation of the outer planet becomes stronger, it affects the
orbital dynamics of the planetesimals more intensely increasing the encounter velocities of those few that enter 
the proto-giant planet's envelope. The combination of small size and high impact velocity causes these planetesimals 
to evaporate more easily. The above increase in encounter velocity occurs for 100 km planetesimals as well. However,
 because of the large sizes of these bodies, the spread in their accreted material stays negligibly small.

\begin{figure}[ht]
\vskip -15pt
\hskip -20pt
\includegraphics[scale=0.36]{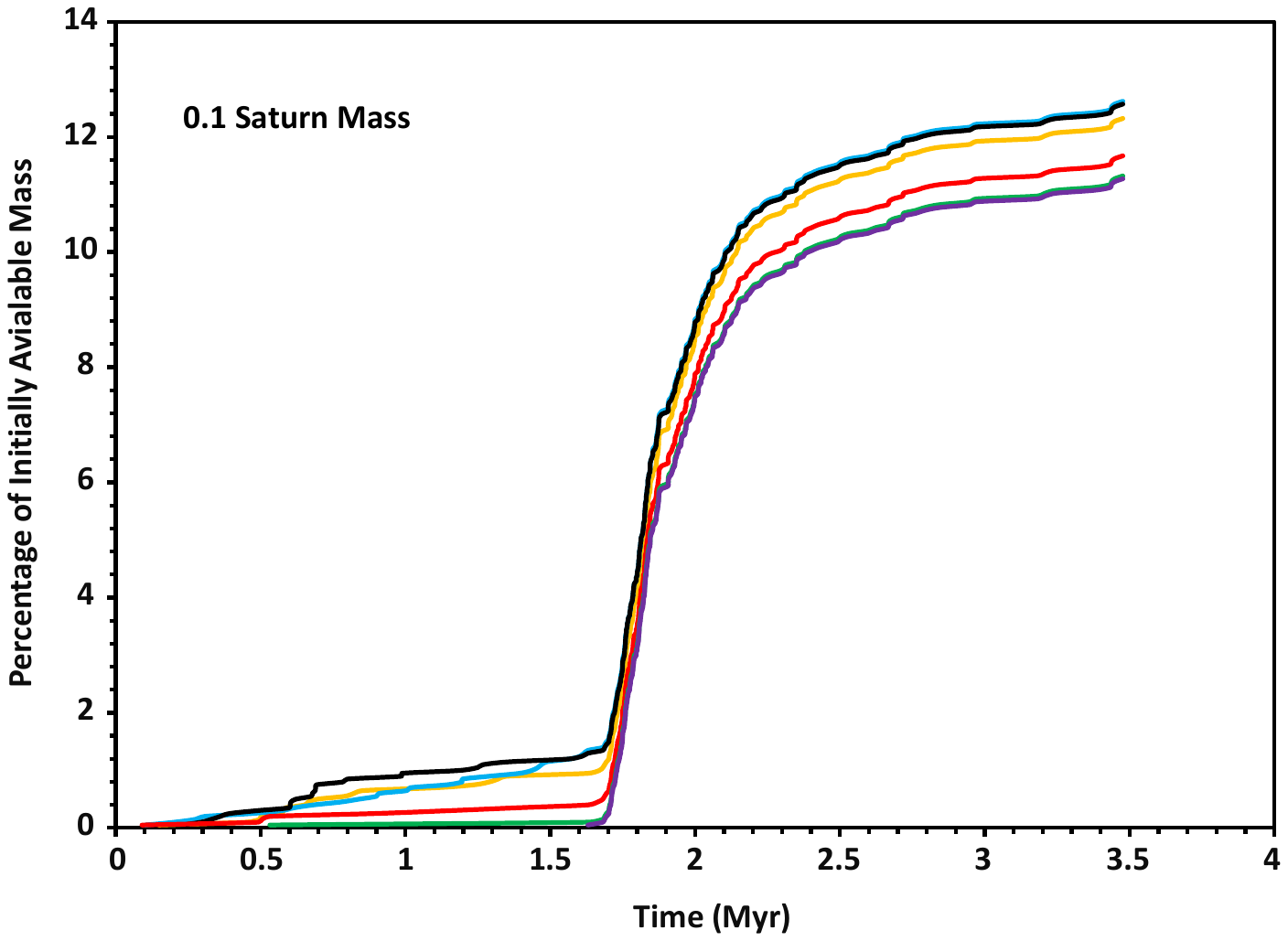}
\vskip -40pt
\hskip -20pt
\includegraphics[scale=0.36]{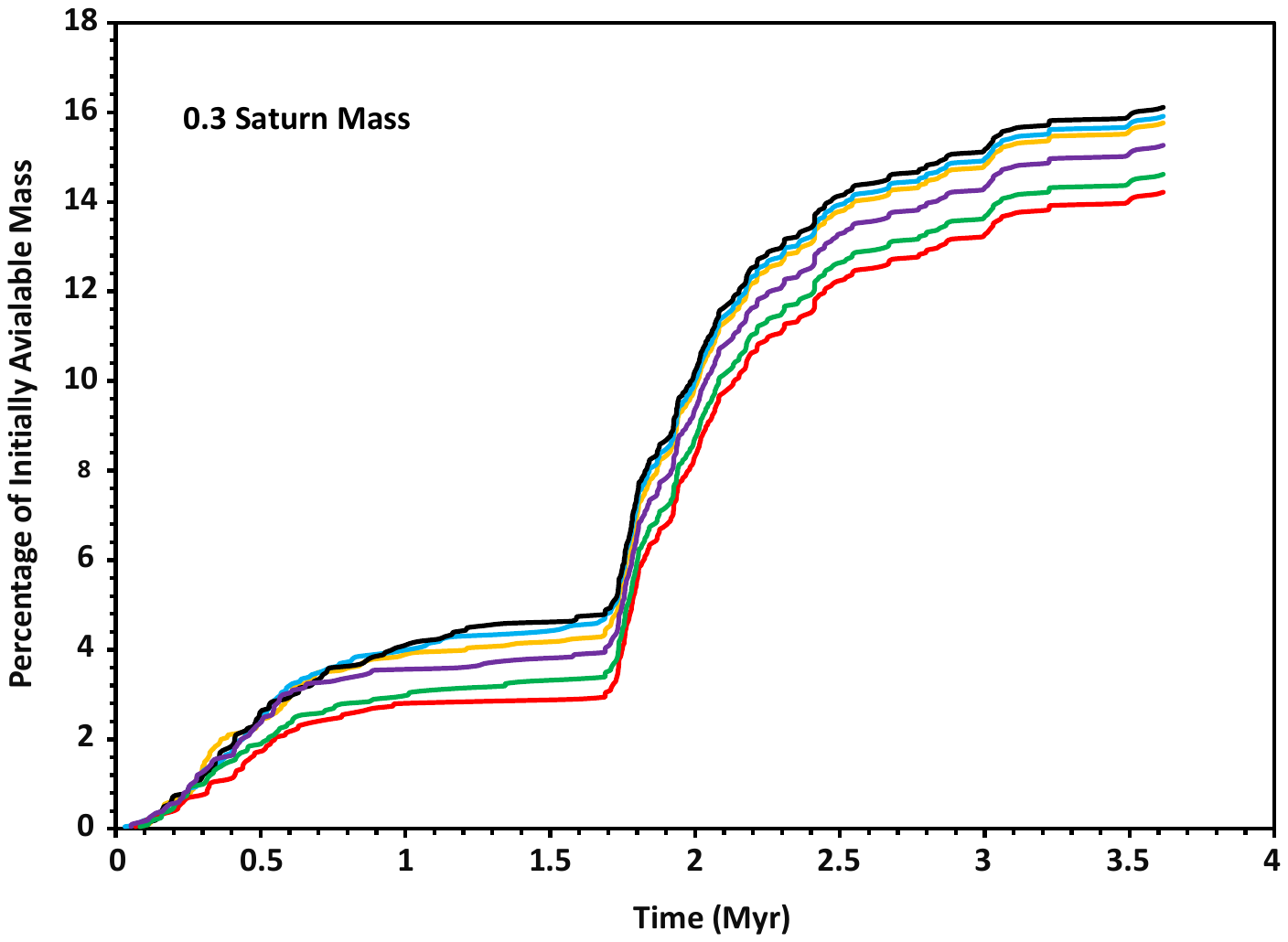}
\vskip -40pt
\hskip -20pt
\includegraphics[scale=0.36]{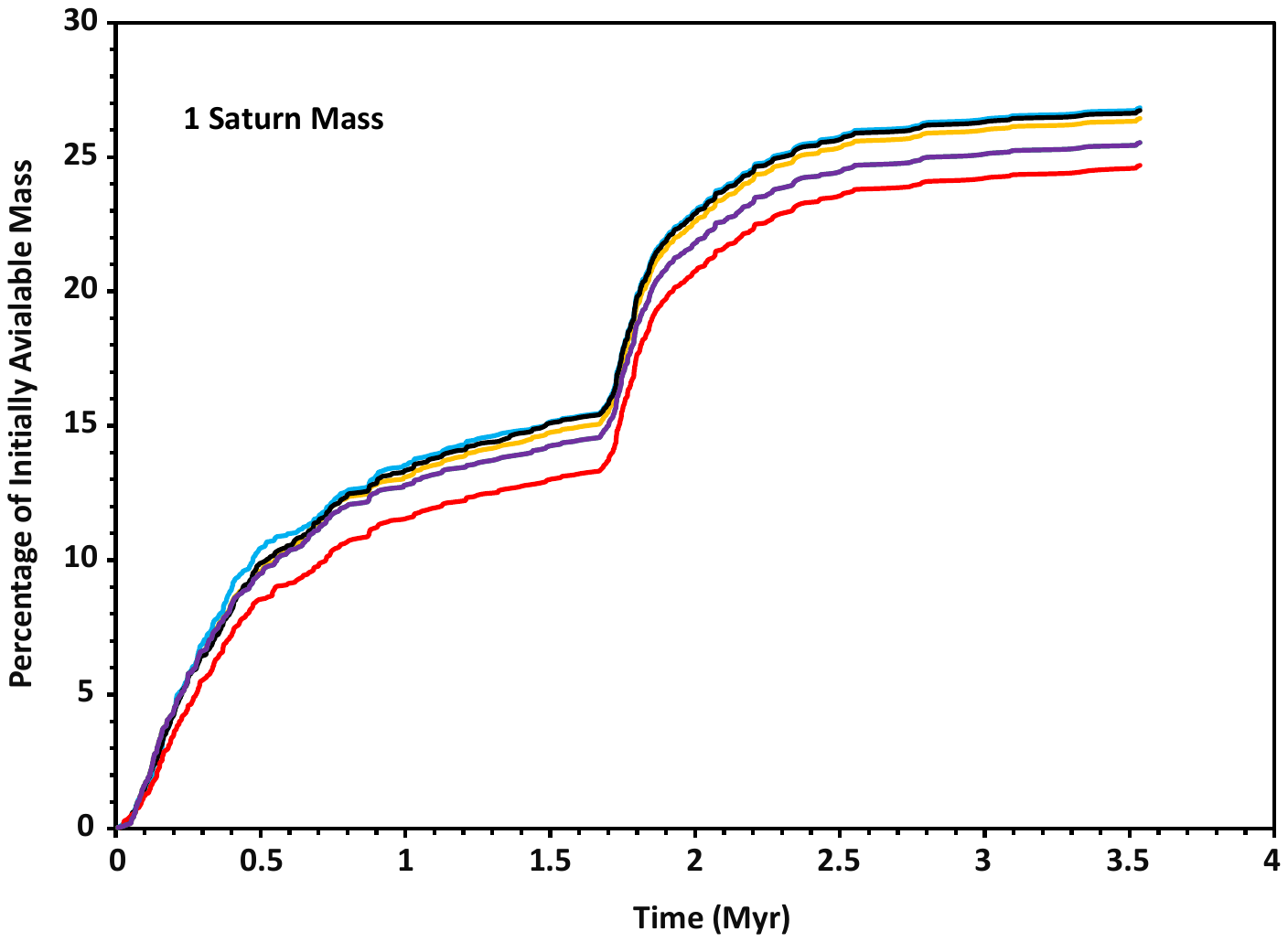}
\vskip -25pt
\caption{Graphs of the mass of the ejected planetesimals as a percentage of the initially available mass
at 8 AU to 12 AU. The color coding is similar to figure 3 corresponding to 1 km and 100 km planetesimals of rock, ice and
mix of ice+rock. From top to bottom each panel corresponds to a different value of the mass of the outer planet.}
\label{fig5}
\end{figure}

\begin{figure}[ht]
\vskip -15pt
\hskip -20pt
\includegraphics[scale=0.36]{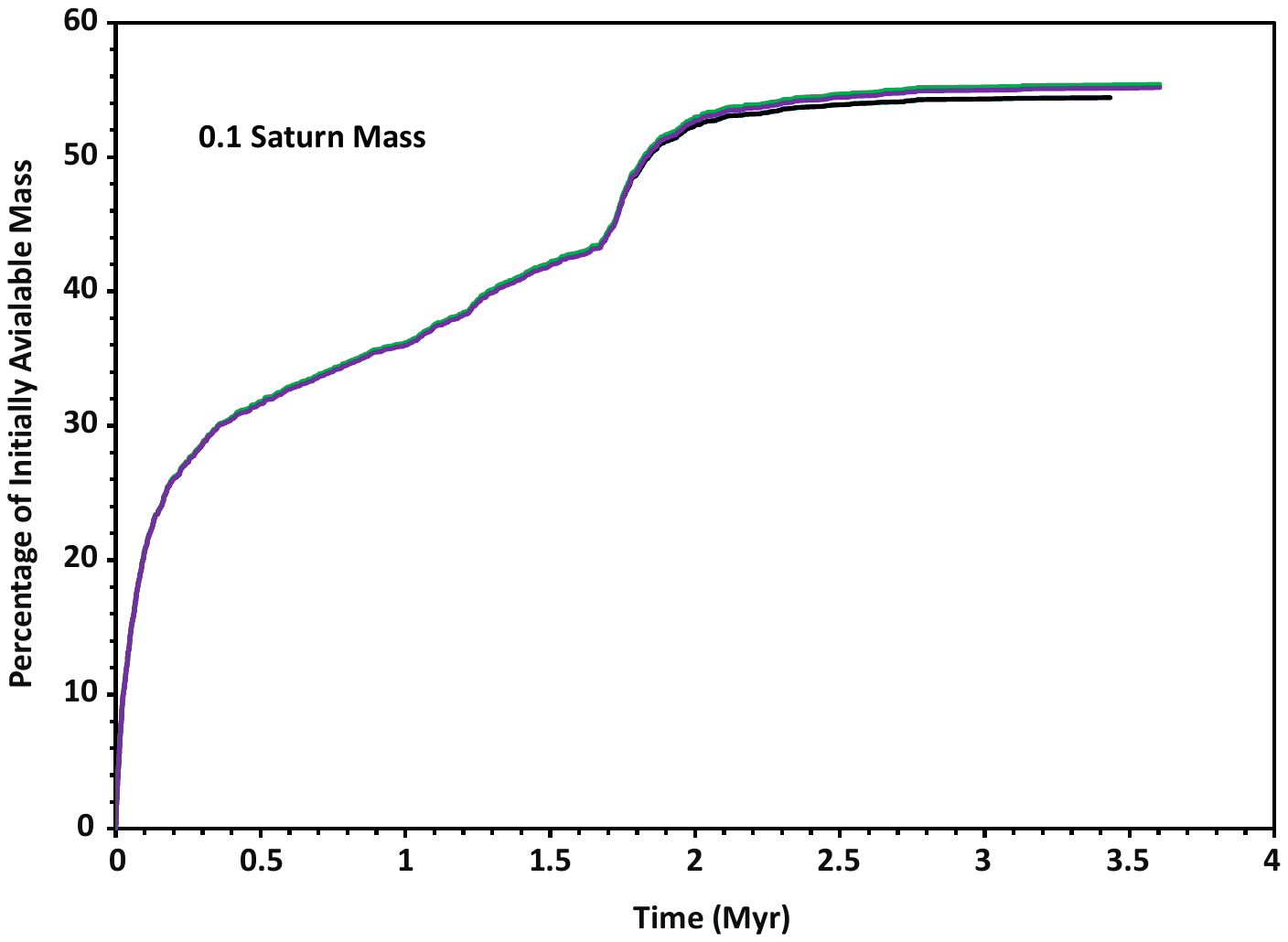}
\vskip -40pt
\hskip -20pt
\includegraphics[scale=0.36]{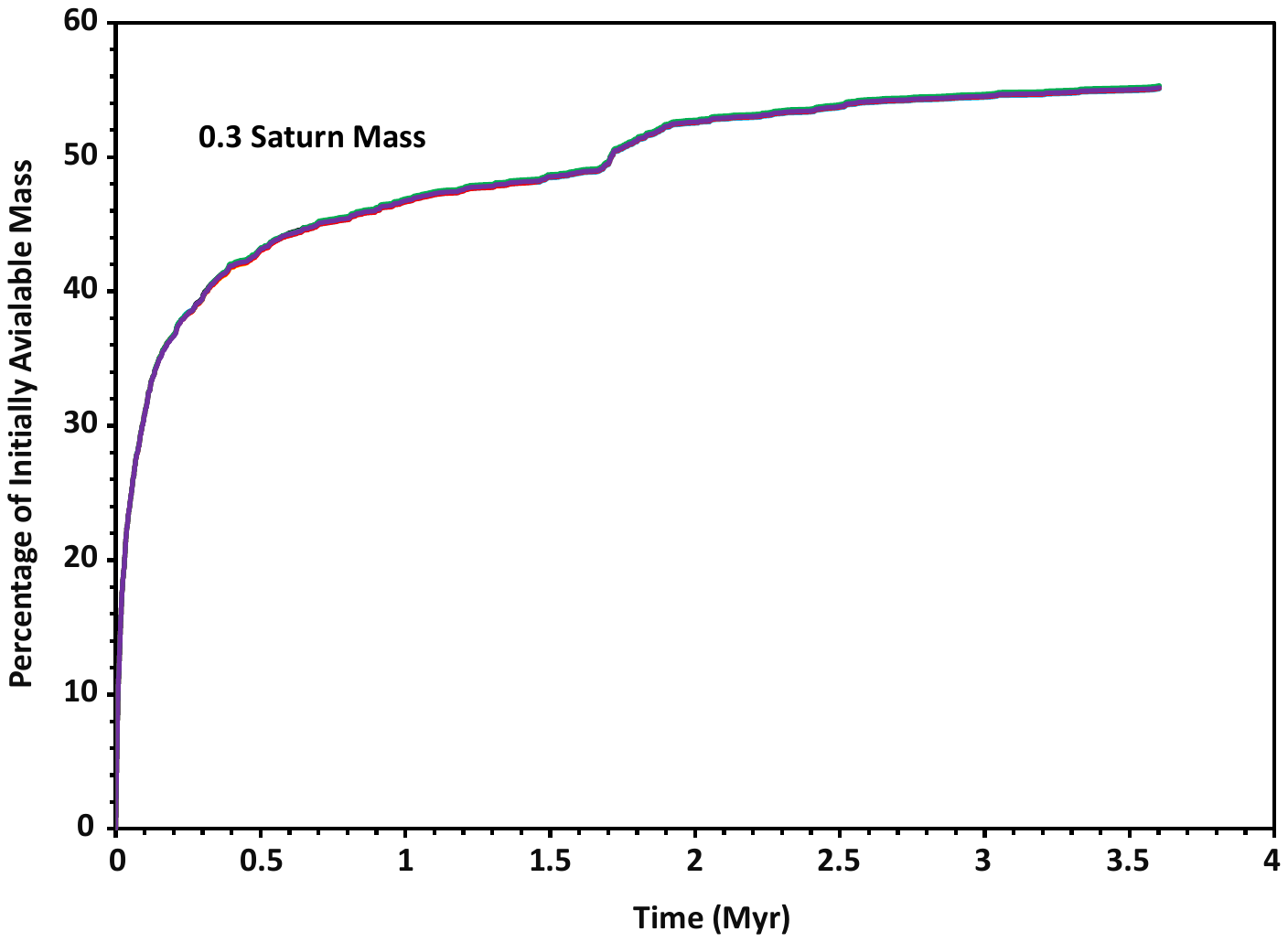}
\vskip -40pt
\hskip -20pt
\includegraphics[scale=0.36]{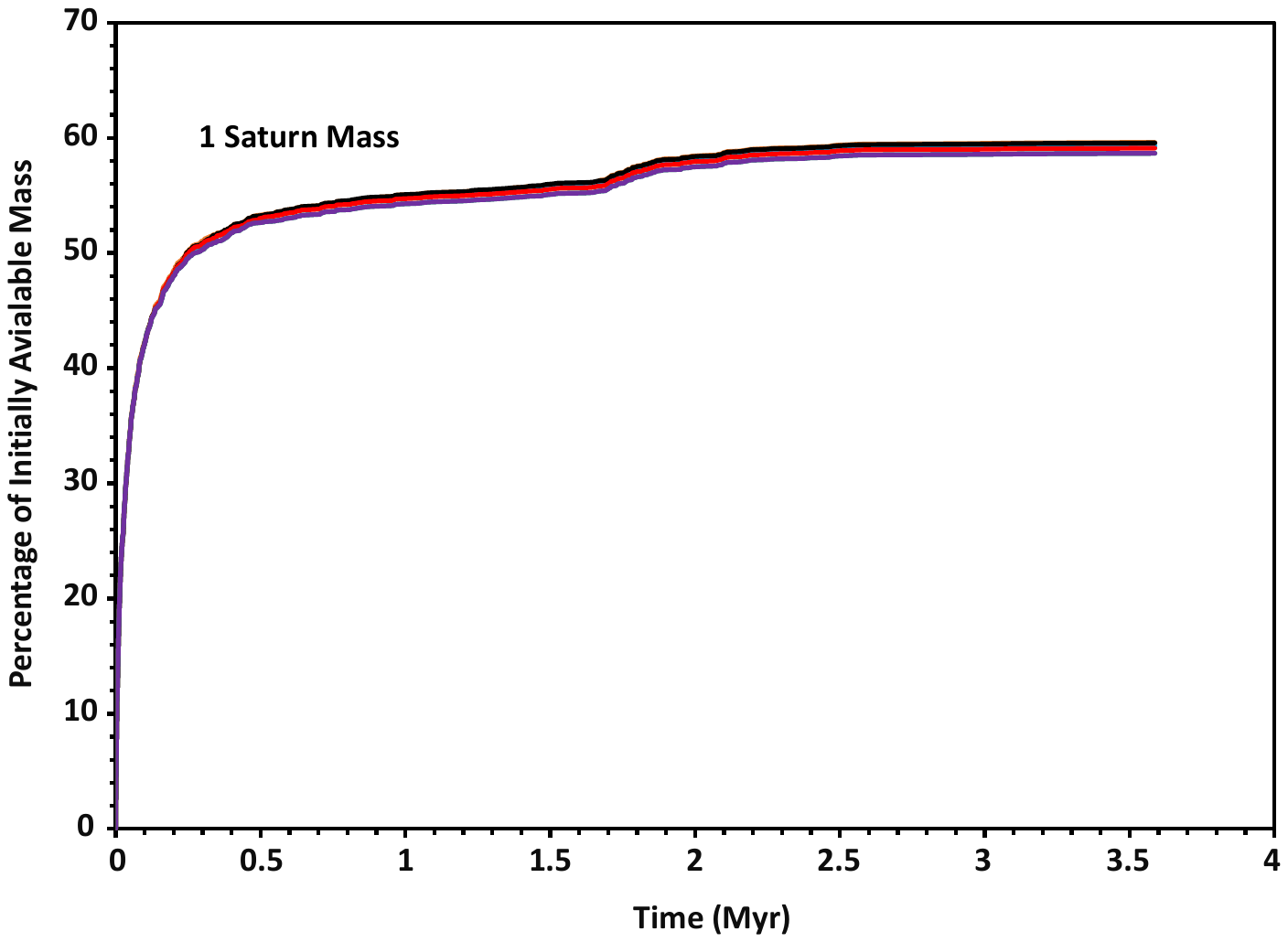}
\vskip -25pt
\caption{Graphs of the planetesimal-mass accreted by the outer planet as a percentage of the initial mass of the 
planetesimals in the region of 8 AU to 12 AU. The color coding is similar to figure 3.}
\label{fig6}
\end{figure}

\begin{figure}[ht]
\vskip -15pt
\hskip -30pt
\includegraphics[scale=0.41]{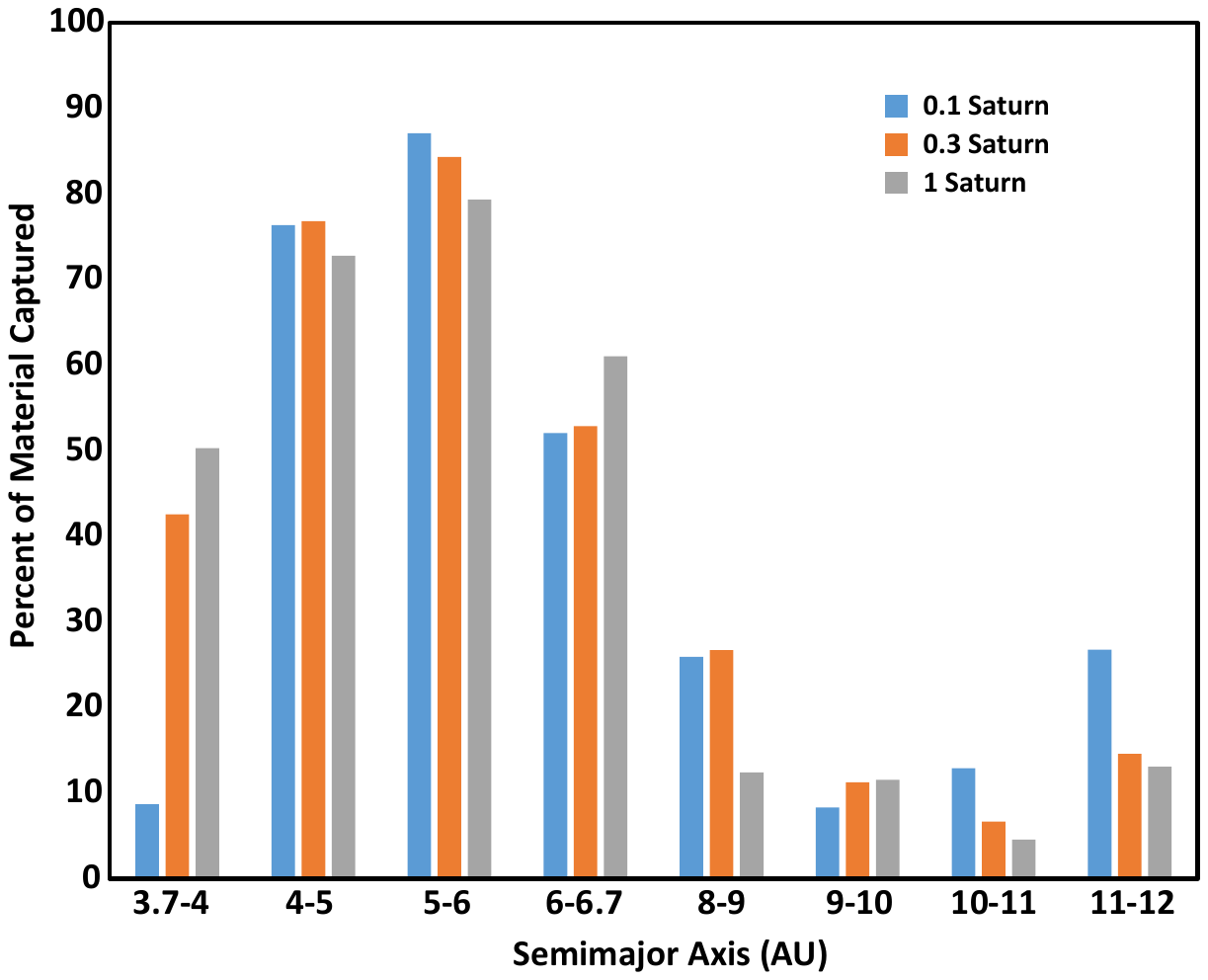}
\vskip -30pt
\caption{Graphs of the percentage of all captured planetesimals for different values of the mass of the outer planet, 
in terms of their initial semimajor axes.}
\label{fig1}
\end{figure}

As demonstrated in papers I and II, when outside the envelope, the orbits of the planetesimals are disturbed by
the perturbation of the proto-giant planet as well. As this planet grows, it scatters planetesimals to farther distances.
In our systems, some of these planetesimals were back-scattered and collided with the outer planet. Figure 6 shows this
for all cases of figure 3. As expected, the rapid increase in the mass of the proto-giant planet during the collapse
of the envelope strengthens the perturbation of this planet causing more planetesimals to collide with the outer planet. 
As a result, the graph of figure 6 shows similar trend as in figure 3. We would like to note that,
as the focus of our study is on determining the contribution of outer planetesimals to the growth and metallicity of the
proto-giant planet, we do not consider any contribution from these planetsimals to the metallicity and growth of the outer 
planet. However, it is important to note that even if the increase in the mass of the outer planet is taken into account,
as shown by figures 3-6, it will tend to enhance the rate of the scattering of planetesimals to farther distances and reduces 
the rates of their encounter with the proto-giant planet's envelope. In other words, the results presented here represent
the best-case scenario.

\section{Implications of the results}

As mentioned earlier, the purpose of this study is to determine the contributions of planetesimals from the 
regions around and beyond the orbit of the outer planet ($8-12$ AU). Figure 7 shows this for different values of the mass 
of this planet and for all compositions and sizes of the planetesimals considered in this study. The bar diagrams in this 
figure represent the total percentage of the combined population of 1 km ice, 1 km rock, 10 km mix, 100 km ice and 100 km 
rock planetesimals that were captured by the envelope of the proto-giant planet in terms of their initial locations. For 
the purpose of comparison, the figures also show the results from Paper II where the system included an outer planet and 
planetesimals were distributed between 3.7 AU and 6.7 AU. 

An inspection of figure 7 indicates that, in general, planetesimals in the region of $8-12$ AU do not have much contribution, 
and most of the contributions come from the planetesimals in the range of $4-7$ AU. 
While, for those planetesimals around 9.5 AU, this is an expected result as the large
majority of these bodies collide with the outer planet (hence the region of $9-11$ AU having the least contribution), 
the small contributions of those outside the 9.5 AU is due to the fact that the perturbation 
of the outer planet scatters many of these bodies out of the system (see figure 2). 
Because this scattering event is random, a small fraction of the scattered planetesimals reach the accretion zone of 
the growing giant planet as well and contribute to its growth. Also, because a more massive outer planet scatters more 
planetesimals and more effectively, as shown by the figure, the amount of the contribution from the region of $8-12$ AU 
decreases by increasing the mass of this planet. We refer the reader to figure 2 where a comparison between the final eccentricities 
of the planetesimals in the two panels clearly indicates that for a low-mass outer planet (upper panel), more planetesimals 
remain in the system. The eccentric orbits of these objects cause some of them to be intercepted and, subsequently, captured
by the proto-giant planet.

Although the focus of this study is on the contributions of planetesimals close to and exterior to the outer planet, 
figure 7 shows another important result that cannot be left un-noted. As shown by this figure,
the largest contributions correspond to the planetesimals from the region of $4-6$ AU. While this is an 
expected result as these planetesimals are close to the accretion zone of the growing giant planet, a deeper look at this 
figure indicates that the level of the contribution of these bodies decreases by increasing the mass of the outer planet. The latter 
firmly confirms the vital role that a planet in the orbit of Saturn plays in the growth and composition of Jupiter.
For the region of $4-6$ AU, the combined perturbation of the outer planet and the growing proto-Jupiter scatters planetesimals 
outside the accretion zone. A more massive planet scatters more planetesimals, reducing the amount of their contributions. 
However, for the regions $3.7-4$ AU and $6-6.7$ AU, this perturbation affects the planetesimals
differently. The same scattering event, along with the increase in the capture cross-section of the growing giant planet, now 
introduces planetesimals to the planet's accretion zone such that the combination
of a more massive outer planet and more massive proto-Jupiter scatters more planetesimals, enhancing their contributions.

\section{Summary and Concluding remarks}

Continuing our efforts on improving the mechanics and accuracy of the core-accretion model, we have investigated the degree of
the contributions of planetesimals from the regions outside the orbit of Saturn to the growth and composition of a growing 
Jupiter-planet. To determine the connection between the magnitude of the contributions and the location of the planetesimals 
in the protoplanetary disk, we used our special purpose integrator ESSTI and integrated the orbits of a large ensemble 
of these objects with semimajor axes ranging 
from 8 AU to 12 AU. To ensure that the calculated contributions would correspond to the complete evolution of the proto-Jupiter,
integrations were carried out for approximately twice the time that of the envelope's collapse. Results pointed to the 
following conclusions:

i) Planetesimals from the regions around and exterior to the outer planet do not show considerable contributions to
the growth and composition of the proto-giant planet. That is due to the fact that the perturbation of the outer
planet scatters many of the surrounding planetsimals to outside the system. The larger the mass of this planet, 
the stronger its scattering effect, and, therefore, smaller contributions. 

ii) Results also show that the rate of planetesimal accretion follows the same timescale as the time of the evolution of 
the envelope. The largest amount of accretion occurs during the envelope's collapse (at 1.7 Myr) when the increase 
the in the local density of the envelope's gas enhances the evaporation of planetesimals and, consequently, the rate 
of their accretion. This rapid growth of the proto-giant planet at around 1.7 Myr is reflected not only in rapid increase 
in its mass accretion, but also in both the rate of the mass ejected from the system and the rate of planetesimals 
captured by the outer planet, in particular when this planet is small.

iii) The above trends are independent of the size and composition of planetesimals, and depend solely on the mass of 
the outer planet. This is an expected result that has to do with the fact that the effects studied here are purely
dynamical, affecting the orbital evolution of the bodies. To determine the contributions of planetesimals to the 
mass and metallicity of the growing giant planet, the chemical structure of the initial protoplanetary disk has
to be take into considerations. The latter is the subject of forthcoming publications.

We would like to note that, to maintain focus on the contributions of exterior planetesimals and the perturbing effect
of the outer planet, we did not include the effect of the nebular gas. It is understood that the drag force of the
nebula will affect the orbital evolution of planetesimals as it causes them to drift toward the central star 
(see section 4 in Paper I). While the latter
is not expected to change the evaporation significantly, as demonstrated in Paper I, it will have 
noticeable consequences on the contribution of those planetesimals that encountered the growing giant planet. How the decrease in
the mass of the colliding bodies affects the amount and type of the mass deposited in the proto-giant planet's envelope
is the subject of future studies.

To conclude, our results strongly imply that in a more realistic simulation, where the contribution
of planetesimals to the growth of the outer planet is also included, as this
planet accumulates mass, fewer planetesimals from the regions outside the influence zone of the
proto-giant planet contribute to its growth and metallicity. As a result, the faster Saturn grows, the less Jupiter’s 
composition will be influenced by the size and composition of the accreted planetesimals from exterior regions meaning
that the mass and composition of Jupiter is mainly and primarily driven by the physical and chemical characteristics
of planetesimals in its vicinity.

\section*{Acknowledgments}
Support from NASA grants 80NSSC18K0519, 80NSSC21K1050, 80NSSC23K0270, and NSF grant AST-2109285 for NH is acknowledged.
We would like to thank the anonymous referee for critically reading our manuscript and their constructive suggestions
and recommendations.

\end{document}